\documentclass{sig-alternate-05-2015}

\usepackage{graphicx}

\usepackage{amsmath}
\usepackage{amssymb}
\usepackage{stackengine}
\usepackage{mathtools}
\DeclareGraphicsExtensions{.eps}
\usepackage{hyperref}
\usepackage[ruled,algonl,boxed]{algorithm2e}
\usepackage{algorithmic}

\newcommand{\norm}[1]{\left\lVert#1\right\rVert}

\begin{document}

\title{Finding Influential Institutions in Bibliographic Information Networks}

\numberofauthors{2}
\author{
\alignauthor
Anubhav Gupta\\
	   \affaddr{Department of Computer Science}\\
       \affaddr{and Automation}\\
       \affaddr{Indian Institute of Science}\\
       \affaddr{Bangalore, India}\\
       \email{anubhav.gupta@csa.iisc.ernet.in}
\alignauthor
M. Narasimha Murty\\
	   \affaddr{Department of Computer Science}\\
       \affaddr{and Automation}\\
       \affaddr{Indian Institute of Science}\\
       \affaddr{Bangalore, India}\\
       \email{mnm@csa.iisc.ernet.in}
}
%
%
%
%
%

\maketitle
\begin{abstract}
Ranking in bibliographic information networks is a widely studied problem due to its many applications such as advertisement industry, funding, search engines, etc. Most of the existing works on ranking in bibliographic information network are based on ranking of research papers and their authors. But the bibliographic information network can be used for solving other important problems as well. The KDD Cup $2016$ competition considers one such problem, which is to measure the impact of research institutions, i.e. to perform ranking of research institutions. The competition took place in three phases. In this paper, we discuss our solutions for ranking institutions in each phase. We participated under team name ``anu@TASL" and our solutions achieved the average NDCG@$20$ score of $0.7483$, ranking in eleventh place in the contest.
\end{abstract}


\section{Introduction}
In recent years, ranking in information network
has attracted a lot of attention due to its wide applications in search engines, advertisements, etc. Most of the work on ranking in information networks is focused on analyzing homogeneous network \cite{page1999pagerank,kleinberg1999authoritative}, which covers only a small part of the large information network. This information network forms very large heterogeneous networks, that are composed of multiple types of objects connected by links between them. Some of the recent work \cite{zhou2007co,sun2009rankclus,meng2013discovering} also takes into consideration, the heterogeneous structure to perform ranking.

Because of the importance associated with the problem of ranking, KDD Cup $2016$ is also based on this exciting problem. The task of KDD Cup Data Mining Contest 2016 hosted by Microsoft Azure is finding influential institutions in the academic network. To be exact, given any upcoming conference such as KDD, the task is to perform ranking of institutions based on predicting how many of their research papers will be accepted in KDD in $2016$. The contestants were given the choice to use any publicly available dataset, for predicting next year's top institutions. We have used Microsoft Academic Graph dataset for this purpose. We have considered this problem as a supervised learning problem and also used temporal information in our ranking model. The competition took place in three phases. In each phase, the contestants were evaluated on one conference which was chosen from some given conferences in that phase. In this paper, we present our solution towards solving this problem. After each phase, we modified our model and used a different algorithm for performing the ranking. In this paper, we describe the algorithms used in each phase individually and demonstrate their performance in all the three phases.

To evaluate the rankings produced by our model, the metric NDCG (Normalized Discounted Cumulative Gain) is used. In information retrieval, Normalized Discounted Cumulative Gain \cite{jarvelin2000ir} is a standard metric for evaluating rankings. The Discounted Cumulative Gain (DCG) at position $n$ is calculated using the following formula:
$$\text{DCG}@n = \sum_{i=1}^{n}\frac{rel_i}{log_2(i+1)}$$
Then, NDCG at position $n$ is defined as:
$$\text{NDCG}@n = \frac{\text{DCG}@n}{\text{IDCG}@n}$$
where $i$ is the predicted rank of an institution and $rel_i$ is its true relevance score. For a perfect ranking algorithm, IDCG is equal to DCG producing an NDCG of $1.0$. For the calculation of true relevance scores, the number of accepted full research papers from every institution is used.

The rest of the papers is organized as follows. Section \ref{sec_notation} describes relevant notation and assumptions made by our approach. Our ranking frameworks are proposed in detail in section \ref{sec_framework}. In section \ref{sec_results}, we perform some experiments and discuss the results of our methods. Finally, we conclude our work in section \ref{sec_conclusion}.

\section{Notation and Assumptions}
\label{sec_notation}
The competition took place in three phases. In each phase, contestants were asked to predict ranking of institutions for a different set of conferences. Let the number of institutions for which we predict rankings be $m$. In the competition, $m=741$. Let us call these institutions $I_1, I_2, ..., I_{m}$. The list of conferences for each of these phases is given in table \ref{tab_phases}.
\begin{table}[tb!]
\centering
\resizebox{6cm}{!}{%
\renewcommand{\arraystretch}{1.3}
\begin{tabular}{|c|l|}
\hline
\textbf{Phase} & \textbf{~~~~~~~~Conferences} \\
\hline \hline
$1$ & SIGIR, SIGMOD, SIGCOMM \\
\hline
$2$ & KDD, ICML \\
\hline
$3$ & FSE, MobiCom, MM \\
\hline
\end{tabular}
}
\caption{List of different conferences for different phases of KDD Cup 2016}
\label{tab_phases}
\end{table}

Following assumptions were imposed by the KDD Cup organizers for finding the relevance of institutions:
\begin{itemize}
    \item For any conference, all the accepted papers are equally important. i.e. each accepted paper has equal vote.
    \item Whenever a research paper is authored by multiple authors, each author is assumed to have made equal contribution to the paper.
    \item If an author is affiliated with multiple institutions, each institution has equal contribution to the paper.
\end{itemize}

\section{Proposed Methods}
\label{sec_framework}
In this section, we discuss our approaches in each phase of the competition.

\subsection{Solution for Phase-1}
In phase-$1$, contestants were asked to predict institution rankings for SIGIR, SIGMOD and SIGCOMM. In this phase, we tried a simple approach where we predict the rankings score of institution for a conference based on their ranking scores in the same conference in previous years. But here, the ranking scores for previous years are weighted differently for different years. Our intuition is that rank of an institution in a conference in $2016$ will depend on its rank in the same conference in year $2015$ more than it depends on its rank in year $2014$ and so on.

\subsubsection{Learning weights of past ranking scores}
\label{sec_weights}
As discussed above, ranking score (fraction of research papers) of an institution in the conference $C_b$ in year $2016$ is obtained by taking the weighted average of ranking scores of this institution in conference $C_b$ in years $2015$, $2014$, etc. The weights for rankings scores of years $2015$, $2014$, etc. are learned using Brown's simple exponential smoothing.

\noindent \textbf{Brown's Simple Exponential Smoothing: }Brown's simple exponential smoothing (exponentially weighted moving average) \cite{smoothing_duke,smoothing_wiki} is a popular technique for smoothing time series data. This technique has the property that it doesn't treat all the past observations equally and assigns them different weights, such that the most recent observation gets more weight than $2^{nd}$ most recent and $2^{nd}$ most recent gets more weight than $3^{rd}$ most recent, and so on. In the simple exponential smoothing method, forecast for a variable $Y$ at time $t+1$ is given by
$$ Y_{t+1} = \alpha [Y_t + (1-\alpha)Y_{t-1} + (1-\alpha)^2Y_{t-2}+...] $$
where $\alpha$ is called the \textit{smoothing constant} and takes value in the range $[0,1]$.

In our case, let $Y_i$'s denote the ranking scores of an institution in different years. We also assume that ranking score of an institution in any year depends only on its ranking scores in previous four years. So, ranking score of an institution in year $2016$ is given by
$$ Y_{2016} = \alpha [Y_{2015} + (1-\alpha) Y_{2014} + (1-\alpha)^2 Y_{2013} + (1-\alpha)^3 Y_{2012}] $$

Now, since $\alpha$ is same for all institutions given a conference, for the purpose of ranking, we only need to learn the parameter $\beta$ such that
$$ Y_{2016} = Y_{2015} + \beta~Y_{2014} + \beta^2~Y_{2013} + \beta^3~Y_{2012} $$

To learn $\beta$ for a conference $C_b$, we use the data upto year $2014$ for training and find the parameter $\beta$ that optimizes the NDCG@$20$ score for ranking of institutions for conference $C_b$ in year $2015$. Then, this $\beta$ is used to predict the ranking of institutions for year $2016$.

The algorithm for ranking institutions is summarized in Algorithm \ref{algo_phase1}

\renewcommand{\algorithmicrequire}{\textbf{Input:}}
\renewcommand{\algorithmicensure}{\textbf{Result:}}

\begin{algorithm}[tb!]
\begin{algorithmic}[1]
\REQUIRE{Bibliographic Information Network $\mathcal{G}$ and \\conference $C_b$ to predict rankings for}
\ENSURE{Rankings of institutions corresponding to the \\conference $C_b$}
\STATE \textbf{Initialize} $all\_w \leftarrow \left[\frac{1}{20}, \frac{2}{20}, ..., 1\right]$
\STATE \textbf{Initialize} max\_score $\leftarrow 0$
\STATE \textbf{Initialize} w\_opt $\leftarrow 1$
\FOR{$k = 1$ to $m$}
  \STATE $r_{b,k}^y \leftarrow$ ranking score (fraction of papers) of institute $I_k$ in year $y$ with respect to conference $C_b$
\ENDFOR
\\ \textbf{// Use ranking scores of institutions in year $2015$ to learn $\mathbf{\beta}$'s}
\FOR{$w$ in $all\_w$}
  \FOR{$k = 1$ to $m$}
    \STATE $s_{b,k}^{2015} \leftarrow r_{b,k}^{2014} + w\cdot r_{b,k}^{2013} + w^2\cdot r_{b,k}^{2012} + w^3\cdot r_{b,k}^{2011}$
  \ENDFOR
  \STATE score $\leftarrow$ NDCG@$20$ score of ranking $s_b^{2015}$, with the true ranking given by $r_b^{2015}$
  \IF{score $\geq$ max\_score}
    \STATE max\_score $\leftarrow$ score
    \STATE $w\_opt \leftarrow w$
  \ENDIF
\ENDFOR
\FOR{$k = 1$ to $m$}
    \STATE $s_{b,k}^{2016} \leftarrow r_{b,k}^{2015} + w\cdot r_{b,k}^{2014} + w^2\cdot r_{b,k}^{2013} + w^3\cdot r_{b,k}^{2012}$
\ENDFOR
\STATE Normalize $s_{b,k}^{2016}$ scores so that they sum to $1$
\STATE Output $s_{b,k}^{2016}$ as relevance scores of institutions
\end{algorithmic}
\caption{\textbf{RankIns$1$}}
\label{algo_phase1}
\end{algorithm}

\subsection{Solution for Phase-2}
In phase-$2$, contestants were asked to predict rankings for the conferences KDD and ICML. The algorithms used for ranking institutions in this phase is same as the one used in phase-$1$, which is given by Algorithm \ref{algo_phase1}.

\subsection{Solution for Phase-3}
In the third phase, contestants were asked to predict ranking of institutions for the conferences FSE, MobiCom and MM. In this section, we discuss the algorithm used in the phase-$3$ of the competition -- \textbf{RankIns$\mathbf{3}$}. This algorithm is different from the algorithm used in previous two phases. We are given the name of a conference, say $C_b$, as a query, and our task is to predict the ranking of institutions based on predicting how many of their research papers will be accepted in this conference in year $2016$. For this purpose, we have used the information provided by heterogeneous bibliographic network upto year $2015$ for training the ranking model.

\subsubsection{Intuition Behind our Approach}
Our ranking model is based on the intuition that rank of the institution $I_k$ for a conference $C_b$ depends on ranks of other similar institutions in conference $C_b$ as well as on its rank in conferences similar to $C_b$. Also, rank of $I_k$ in conference $C_b$ depends on rank of $I_k$ in the past instances of $C_b$, i.e. rank of $I_k$ in conference $C_b$ in $2015$, $2014$, etc. Therefore, conferences and institutions are represented as feature vectors as such a representation allows the similarity between institutions and conferences, and between the institutions to be measured in terms of cosine of their feature vectors.

The RankIns$2$ algorithm works as follows. The existing problem is first transformed into a learning-to-rank problem \cite{liu2009learning} and then learning-to-rank techniques such as RankSVM \cite{citeulike:477598}, RankBoost \cite{964285}, AdaRank \cite{1277809} can directly be used to solve the learning problem. To rank institutions, RankIns$2$ first constructs the data matrix for year $2016$ and the data matrices upto year $2015$ are used to train the learning-to-rank model. Then this learned model is used to predict ranks of institutions for data matrix of year $2016$.

\subsubsection{Construction of Features Vectors}
RankIns$2$ uses the following method in order to construct feature vectors of institutions:
\begin{itemize}
  \item Authors and keywords associated with papers are used to identify the institutions.
  \item Instead of using individual authors, clusters of authors are used as features. To perform the clustering of authors, methods such as BAGC \cite{xu2012model} and SI-Cluster \cite{zhou2013social} can be used. Let the number of clusters be $K$. Let these $K$ clusters be represented by their respective centers $a_1, a_2, ..., a_K$.
  \item For the datasets in which topic information is not available, methods such as Latent Dirichlet Allocation \cite{blei2003latent} and Probabilistic Latent Semantic Indexing \cite{hofmann1999probabilistic} can be used to find topics of research papers.\\
\end{itemize}

Feature vectors for all the conferences are constructed in the same way. Let us say, feature vector of institution $I_k$ is given by:
$$
I_{k} = \overbrace{\left[\begin{matrix}
  \alpha_{i_1} & \alpha_{i_2} & .& .& \alpha_{i_K}
 \end{matrix} \right.}^{Author~Clusters}
 \overbrace{\left.\begin{matrix}
  \beta_{i_1} & \beta_{i_2} & .& .& \beta_{i_s}
 \end{matrix} \right]}^{Topics}
$$

\noindent and feature vector of conference $C_b$ is given by:
$$
C_{b} = \overbrace{\left[\begin{matrix}
  \alpha_{c_1} & \alpha_{c_2} & .& .& \alpha_{c_K}
 \end{matrix} \right.}^{Author~Clusters}
 \overbrace{\left.\begin{matrix}
  \beta_{c_1} & \beta_{c_2} & .& .& \beta_{c_s}
 \end{matrix} \right]}^{Topics}
$$

Here, $s$ is the number of distinct topics in the dataset. Now, the feature vector of institution $I_k$ corresponding to conference $C_b$ is obtained by taking the element-wise multiplication of their feature vectors, and is given by\\

$ ~{I_k}^{(b)} = I_k \odot C_b $
$$
 ~{I_k}^{(b)} = \left[ \begin{matrix}
  \alpha_{i_1}\alpha_{c_1} & .& .& \alpha_{i_K}\alpha_{c_K} & \beta_{i_1}\beta_{c_1} & .& .& \beta_{i_s}\beta_{c_s}
 \end{matrix} \right]
$$

Let these feature vectors corresponding to conference $C_b$ be together denoted by matrix $M_b$, which is defined as follows:

\def\tmp{%
  \begin{matrix}
  \xleftarrow{\hspace*{1cm}}{I_1}^{(b)}\xrightarrow{\hspace*{1cm}}\\
  \xleftarrow{\hspace*{1cm}}{I_2}^{(b)}\xrightarrow{\hspace*{1cm}}\\
 .\\
 .\\
 .\\
  \xleftarrow{\hspace*{1cm}}{I_k}^{(b)}\xrightarrow{\hspace*{1cm}}
 \end{matrix}
}%

$$ M_b = \left( \tmp \right)_{k\times (K+s)} $$

\noindent So, the matrix $M_b$ is a $k\times d$ matrix, where $d = K + s$.
\vspace{10pt}

\subsubsection{Constructing Data Matrix for Year 2016}
\label{sec:data_mat}
Here, we describe the procedure to find data matrix for year $2016$ corresponding to conference $C_b$. Since, the institutions in data matrices are represented in the form of clusters of authors and topics, the data matrix for year $2016$ can be obtained by using data matrices of past years because any new research paper will belong to either some of the existing topics or some mixture of them, and will be written by an author belonging to one of the ranked categories. Let us assume that data matrix of any year depends only on the data matrices of previous three years. So, the data matrix of $2016$ depends only on data matrices of years $2015$, $2014$ and $2013$. Let us denote by $M_b^{(y)}$, data matrix for year $y$ corresponding to conference $C_b$.

To learn the data matrix for year $2016$, we make the assumption that data matrix of any year $y$ is a linear combination of data matrices of years $y-1$, $y-2$ and $y-3$. i.e. data matrix of year $2016$

\begin{equation}
\hat{M}_b^{(2016)} = w_1\cdot M_b^{2015} + w_2\cdot M_b^{2014} + w_3\cdot M_b^{2013}
\end{equation}

\noindent where $w_1$, $w_2$ and $w_3$ are the weights of data matrices for years $2015$, $2014$ and $2013$ respectively. Let us represent these weights by a $3$-dimensional vector $\mathbf{w} = [w_1~w_2~w_3]^T$. To learn these weights, we first learn the initial weight vector $\textbf{w}^{(0)}$ by solving the following optimization problem using the data upto year $2015$ for training:

  \begin{equation}
    \mathbf{w}^{(0)} = arg\min_{\mathbf{w}} \norm{M_b^{2015} - \hat{M}_b^{2015}}_F^2
  \end{equation} where
  
  $$ \hat{M}_b^{2015} = w_1\cdot M_b^{2014} + w_2\cdot M_b^{2013} + w_3\cdot M_b^{2012} $$
  
To find the solution of the above equation, we introduce some additional notation. Let $n$ be the number of authors in the dataset. Let us denote by $R$, a $(n\times d\times 3)$ matrix tensor, such that
$$
R_{ij:} = \left[
  (M_b^{2014})_{ij}~~(M_b^{2013})_{ij}
  ~~(M_b^{2012})_{ij}
 \right]
$$

Let us define a matrix $X \in \mathbb{R}^{(md)\times 3}$ whose rows are the tube fibers of the tensor $R$, as follows:
$$
X = \left[R_{11:}~~R_{12:}~~.~~.~~.~~R_{md:} \right]^T
$$

Also, let $$\mathbf{z} = 
  \left[
  (M_b^{2015})_{11}~~(M_b^{2015})_{12}~~.
  ~~.~~.~~(M_b^{2015})_{md}
 \right]^T
$$

Then, one can easily see following relationship:
$$ \norm{M_b^{2015} - \hat{M}_b^{2015}}_F^2 = \norm{\mathbf{z} - X\mathbf{w}}^2 $$

Using this equality, the problem given in equation (2) reduces to
$$ \mathbf{w}^{(0)} = arg\min_{\mathbf{w}} \norm{\mathbf{z} - X\mathbf{w}}^2 $$

The above equation has a closed form solution, which is obtained by equating the gradient of the above equation to zero. The solution is given by:
\begin{equation}
  \mathbf{w}^{(0)} = (X^TX)^{-1}X^T\mathbf{z} \\
\end{equation}

After learning the initial weight vector $\mathbf{w}^{(0)}$, the algorithm iteratively updates this weight vector by using the data of previous years. We use data upto years $2014$, $2013$, $...$ at each iteration of the algorithm. At $l^{th}$ iteration, the following optimization problem is solved to update the weight vector $\textbf{w} = [w_1~w_2~w_3]^T$: 
\setlength{\arraycolsep}{0.0em}
\begin{eqnarray}
\mathbf{w}^{(l)}&{}={}&arg\min_{\mathbf{w}} \norm{M_b^{(2015-l)} - \hat{M}_b^{(2015-l)}}_F^2\nonumber\\
&&{+}\: \lambda_l \norm{\textbf{w}^{(l-1)} - \textbf{w}}^2
\end{eqnarray}
\setlength{\arraycolsep}{5pt}
  Here,
  $$ \hat{M}_b^{(2015-l)} = w_1\cdot M_b^{(2014-l)} + w_2\cdot M_b^{(2013-l)} + w_3\cdot M_b^{(2012-l)}$$

The $\lambda_l$'s above are the hyper-parameters, used to make sure that recent links in the network are given more importance than past links. The second term in the above equation is a regularizer and it ensures that the updated weight vector is not too far away from its previous value.

Again, for easy calculation, we introduce some notation. Let us denote by $S^{(l)}$, a $(n\times d\times 3)$ matrix tensor, such that
$$
S^{(l)}_{ij:} = \left[
  (M_b^{(2014-l)})_{ij}~~(M_b^{(2013-l)})_{ij}~~(M_b^{(2012-l)})_{ij}
 \right]
$$

Let us define the matrix $X_l \in \mathbb{R}^{(md)\times 3}$ whose rows are the tube fibers of matrix tensor $S^{(l)}$, as follows:
$$
X_l = \left[S^{(l)}_{11:}~~S^{(l)}_{12:}~~.~~.~~.~~S^{(l)}_{md:} \right]^T
$$

Also, let us define the vector $\mathbf{z}_l$ as follows:
$$\mathbf{z}_l = 
  \left[
  (M_b^{(2015-l)})_{11}~~(M_b^{(2015-l)})_{12}~~.~~.~~.~~(M_b^{(2015-l)})_{md}
 \right]^T
$$

Then, one can observe following relationship:
$$ \norm{M_b^{(2015-l)} - \hat{M}_b^{(2015-l)}}_F^2 = \norm{\mathbf{z_l} - X_l\mathbf{w}}^2 $$

Using the above equality, the problem given by equation (4) reduces to
$$  \mathbf{w}^{(l)} = arg\min_{\mathbf{w}} \norm{\mathbf{z_l} - X_l\mathbf{w}}^2 + \lambda_l \norm{\textbf{w}^{(l-1)} - \textbf{w}}^2 $$

By taking the gradient of above equation and equating it to zero, we get:
$$ 2X^T(X_l\mathbf{w} - \mathbf{z^l}) - 2\lambda_l(\mathbf{w}^{(l-1)} - \mathbf{w})~ =~ \mathbf{0}$$

By, solving the above equation, we obtain the following solution for $\mathbf{w}^{(l)}$
\begin{equation}
\mathbf{w}^{(l)} = (X_l^TX_l + \lambda_l I)^{-1}(X_l^T\mathbf{z_l} + \lambda_l \mathbf{w}^{(l-1)})
\end{equation}

\subsubsection{Solving the learning problem}
Using the weights learned above, we create the data matrix of all institutions for year $2016$ corresponding to conference $C_b$ as given in equation (1). Then, using the data matrices upto year $2015$ for all the conferences for training, we treat the problem as a regression problem and predict the scores for institutions using random forest regression method. These scores are the predicted relevance scores of institutions.

\renewcommand{\algorithmicrequire}{\textbf{Input:}}
\renewcommand{\algorithmicensure}{\textbf{Result:}}

\begin{algorithm}[tb!]
\begin{algorithmic}[1]
\REQUIRE{Bibliographic Information Network $\mathcal{G}$ and \\conference $C_b$ to predict rankings for}
\ENSURE{Rankings of institutions corresponding to conference $C_b$}
\STATE Construct feature vectors for all the institutions and \\all the conferences
\STATE Construct feature vectors corresponding to every \\institution and conference pair
\STATE Create matrices $M_b^{(i)}$ for $i = 2011, 2012, ..., 2015$ for conference $C_b$
\STATE Create matrix $X$ and vector $\mathbf{z}$ as discussed in Section \ref{sec:data_mat}
\STATE Initialize $$\mathbf{w}^{(0)} \leftarrow (X^TX)^{-1}X^T\mathbf{z}$$
\FOR{$l = 1$ to $u$}
\STATE Create matrix $X_{l}$ and vector $\mathbf{z}_l$ as discussed in Section \ref{sec:data_mat}
\STATE Update the weight vector $\mathbf{w}$ as follows:
  $$\mathbf{w}^{(l)} \leftarrow (X_l^TX_l + \lambda_l I)^{-1}(X_l^T\mathbf{z_l} + \lambda_l \mathbf{w}^{(0)})$$
\ENDFOR
\STATE Final weight vector $\mathbf{w^*} \leftarrow \mathbf{w}^{(u)}$
\STATE Construct data matrix for year $y$ as
  $$ \hat{M}_b^{2016} = \mathbf{w}^*_1\cdot M_b^{2015} + \mathbf{w}^*_2\cdot M_b^{2014} + \mathbf{w}^*_3\cdot M_b^{2013} $$
\STATE Train the learning model with data upto year $2015$
\STATE Use this model to make predictions for all the institutions in $\hat{M}^{2016}_b$
\end{algorithmic}
\caption{\textbf{RankIns$2$}}
\label{algo:RankIns2}
\end{algorithm}

Algorithm \ref{algo:RankIns2} summarizes the working of our method. The input to RankIns$2$ is a bibliographic information network $\mathcal{G}$ and a conference $C_b$ to predict rankings for. RankIns2 then predicts the rankings of institutions corresponding to this given conference.

\section{Experiments and Results}
\label{sec_results}
\begin{figure*}[htb!]
    \includegraphics[scale=0.42]{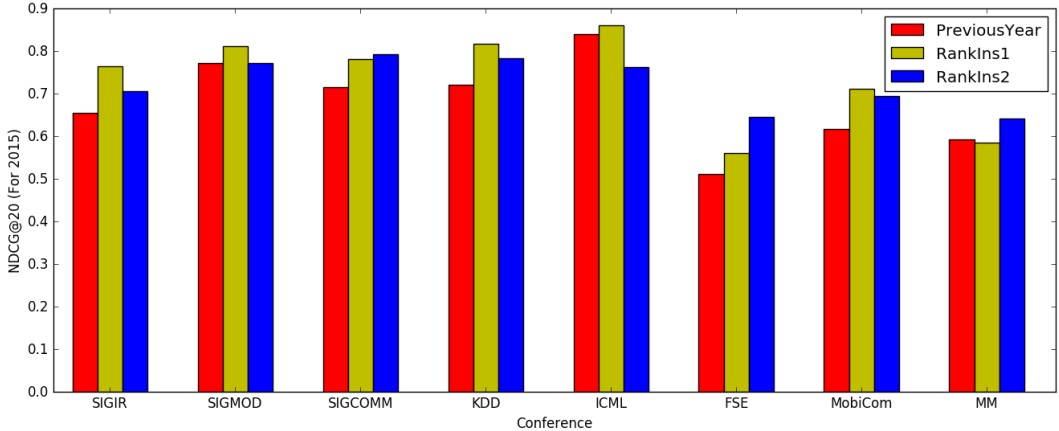}  \caption{Performance of RankIns$1$ and RankIns$2$ on validation dataset}
  \label{fig_res_validation}
\end{figure*}

\subsection{Dataset Description}
\begin{figure}[tb!]
\begin{center}
\includegraphics[scale=0.3]{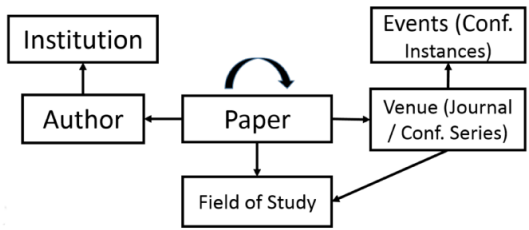}
\caption{Schema of Microsoft academic graph dataset, adapted from \cite{sinha2015overview}}
\label{fig_mag}
\end{center}
\end{figure}

For performing the ranking of institutions, we have used the Microsoft Academic Graph (MAG) dataset which is a huge dataset containing information about scientific documents from different research domains. Figure \ref{fig_mag} shows the schema of MAG.

Along with the MAG dataset, organizers of KDD Cup additionally provided two separate files for the purpose of this competition. These two files contain the names of institutions for which we have to predict the rankings and details of full research papers accepted in the asked conferences during the time period $2011$ - $2015$.

We merged MAG dataset with the given two files and then divided the dataset on a yearwise basis. i.e. we created several small heterogeneous bibliographic networks by considering the research papers of year $2011$, $2012$, ..., $2015$ respectively.

\subsection{Evaluation}
To evaluate the proposed ranking algorithms, the metric \textit{Normalized Discounted Cumulative Gain} is used:\\

\noindent \textbf{Normalized Discounted Cumulative Gain:} In information retrieval, Normalized Discounted Cumulative Gain (NDCG) \cite{jarvelin2000ir} is a standard metric for evaluating rankings. The Discounted Cumulative Gain (DCG) at position $n$ is calculated using the following formula:
$$\text{DCG}@n = \sum_{i=1}^{n}\frac{rel_i}{log_2(i+1)}$$

\noindent Then, NDCG at position $n$ is defined as:
$$\text{NDCG}@n = \frac{\text{DCG}@n}{\text{IDCG}@n}$$
where $i$ is the predicted rank of an institution and $rel_i$ is its true relevance score. For a perfect ranking algorithm, IDCG is equal to DCG producing an NDCG of $1.0$.\\

\subsection{Results and Discussion}
To evaluate the proposed methods, one conference out of all the given conferences was selected in each phase, to form the test dataset. Name of the conference selected and the results of our methods for these conferences are given in Table \ref{tab_res_test}. We also create a validation set, where we predicted the ranks of institutions for all the conferences appearing in the three phases in year $2015$. The performance of RankIns$1$ and RankIns$2$ for each of these conferences is given in figure \ref{fig_res_validation}. To compare our methods, we also take one more ranking of institutions. This ranking, called \textbf{PreviousYear}, ranks an institution corresponding to a conference in year $y$, same as its rank in $y-1$ for this conference.

In our implementation of RankIns$2$, we have used $u=2$ and $\lambda_l=200$. The number of clusters of authors used to create feature vectors is $500$.

The results on validation set show that proposed methods RankIns$1$ and RankIns$2$ outperforms PreviousYear for $7$ our of $8$ conferences. From results, however, it is not clear whether RankIns$1$ should be preferred over RankIns$2$ or not. But RankIns$1$ has a clear drawback compared to RankIns$2$. For the conferences where there is a sudden change in the rankings of institutions in the current year, but rankings are similar in previous years, rankings predicted by RankIns$1$ will be heavily affected by current year rankings (which seems to be an outlier). 

\begin{table}[tb!]
\centering
\renewcommand{\arraystretch}{1.3}
\begin{tabular}{|c|l|c|l|}
\hline
\textbf{Phase} & \textbf{Conference} & \textbf{NDCG@20} & \textbf{Algorithm} \\
\hline \hline
$1$ & SIGIR & $0.7364464224$ & RankIns$1$ \\
\hline
$2$ & KDD & $0.7815387475$ & RankIns$1$ \\
\hline
$3$ & MM & $0.7210261755$ & RankIns$2$ \\
\hline
\end{tabular}
\caption{Performance of RankIns$1$ and RankIns$2$ on test dataset}
\label{tab_res_test}
\end{table}

Data preprocessing was performed on an Intel Xeon E5-2640 v3 2.60 GHz (Haswell-based) machine with 128 GB of memory. All the experiments were performed on an Intel Core I$5$ $4200$U machine with $4$GB of memory. Our codes are written in Python 2.7.11. Our codes are all single-threaded. 

\section{Conclusion}
\label{sec_conclusion}
In this paper, we have discussed the problem of ranking institutions and proposed our methods RankIns$1$ and RankIns$2$. Both RankIns$1$ and RankIns$2$ consider the ranking problem as a supervised learning problem and both the approaches consider the bibliographic data in the form of time-series. RankIns$1$ predicts the ranking scores of institutions by giving weights to their ranking scores in previous years. Whereas, RankIns$2$ works by transforming the problem into a learning-to-rank problem and then uses the learning-to-rank framework to predict the rankings of institutions.

\section{Acknowledgments}
We would like to acknowledge SIGKDD for organizing such a challenging and inspiring competition. We also thank Microsoft for providing the dataset and the user friendly platform for the competition. Finally, we would like to acknowledge Sharad Nandanwar of IISc for the fruitful discussions we had with him.

\bibliographystyle{abbrv}
\bibliography{refs} 
\end{document}